\documentclass[graybox]{svmult}
\pdfoutput=1

\usepackage{mathptmx}       
\usepackage{helvet}         
\usepackage{courier}        
\usepackage{type1cm}        
\usepackage{makeidx}         
\usepackage{graphicx}        
\usepackage{multicol}        
\usepackage[bottom]{footmisc}
\usepackage{amsmath}
\usepackage{upgreek}
\usepackage{natbib}
\usepackage{hyperref}
\usepackage{doi}

\begin{document}

\title*{12\\Reduction of wind power variability through geographic diversity}
\titlerunning{Reduction of wind power variability}
\authorrunning{\textit{Variable Renewable Energy and the Electricity Grid}}

\author{Mark Handschy, Stephen Rose, and Jay Apt }

\institute{This is an accepted manuscript of Chapter 12 from {\em Variable Renewable Energy and the Electricity Grid}, by Jay Apt and Paulina Jaramillo, published by Taylor \& Francis (2014), \doi{10.4324/9781315848709}. }

\maketitle

\abstract*{Each }

\section{Introduction}
\label{sec:1}
The variability of wind-generated electricity can be reduced by aggregating the outputs of wind generation plants spread over a large geographic area. In this chapter we utilize Monte Carlo simulations to investigate upper bounds on the degree of achievable smoothing and clarify how the degree of smoothing depends on the \textit{number} of plants and on the size of the geographic \textit{area} over which they are spread. 

We model two distinct benefits of geographic diversity that have different behaviors: (1) increased tendency of generation level to lie near its mean and (2) decreased tendency for it to lie near its extremes. Gaussian or normal probability distributions give accurate estimates of the first but underestimate the second. The second benefit, which has particular importance for electric grid reliability, has not been widely treated before. 

The effect of geographic diversity on wind generation variability has been investigated by many, starting with \cite{thomas}. Significant early work was carried out by \cite{molly}, \cite{justus}, \cite{kahn}, \cite{farmer}, and \cite{carlin}. Some researchers have focused on particular geographic regions, including the U.S. Midwest \citep{archer,archer2,fisher} and the Nordic countries \citep{holt}. Others have investigated the effect on the frequency spectrum of the generated power \citep{mcnerney,beyer,nan,katz,tarroja}. \cite{hasche} has modeled how the smoothing benefit saturates as the number of generator sites within a region increases. Some recent investigations have focused on the effects of spreading arrays of wind generators over especially large distances. \cite{kempton} and \cite{dvorak} considered an array of wind farms distributed along the entire extent of the U.S. East Coast, while \cite{fertig} and \cite{louie} evaluated the smoothing effect on wind generation by interconnections between independent system operators (ISOs) across the U.S., and \cite{huang} considered wind farms spread over the Great Plains of the U.S. from Montana to Texas. These previous studies largely model the generation of an array of hypothetical wind plants by simulating the output of each individual plant from a few-year historical weather record, with some additionally fitting the empirical distribution of modeled array generation levels to some chosen standard parametric probability distribution. 

\begin{figure}[t]
\includegraphics[width=11.7cm]{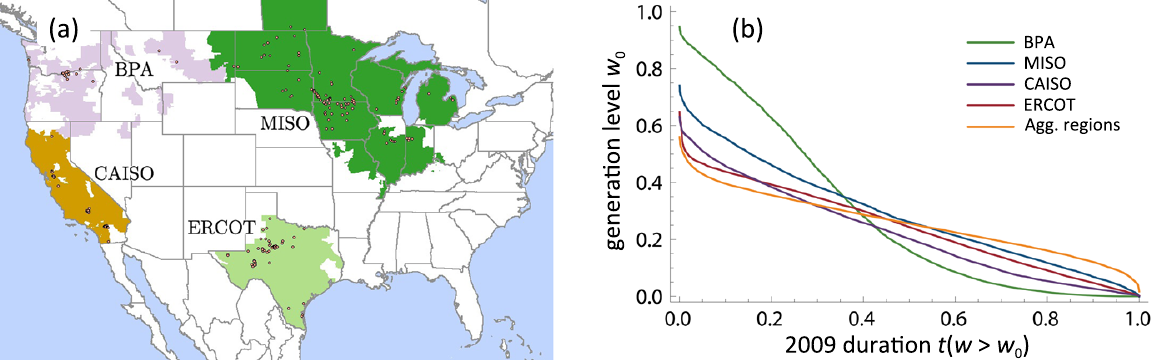}
\caption{Wind regions (a) and wind power duration curves (b) for the year 2009, after \cite{fertig}.}
\label{emily}
\end{figure}

As an example of that work, consider the four U.S. regions examined by \cite{fertig}. Figure \ref{emily} shows the regions and a generation duration curve for each region and for a sum of all regions using hourly data. The sum simulates the generation duration curve for the year 2009 if all four regions had been interconnected. The duration curve for the Bonneville Power Administration (BPA) region shows the effect of high correlation between turbines that are mostly located in the Columbia River gorge. The data shown in Figure \ref{emily}(b) encompass only a single year, and during that interval there is no hour where the sum of the power from all four regions drops to zero. If data were available for a long enough interval, perhaps many decades, we would expect that in at least one hour there would be no wind generation at all; the duration curves would always go to zero at 100 percent of the hours. 

For probabilistic consideration of long-term capacity planning, the interesting question is: for a given percent of hours (e.g.\ 99.97\%), what is the minimum power output that can counted upon from widely-distributed wind turbines?

\section{Monte Carlo model }
\label{sec:2}
To gain better insight into the implications of geographic diversity for low-probability events that influence grid reliability, and to more clearly articulate the dependence of the different smoothing benefits on the number of included wind plants, we created a Monte Carlo model of an array with an adjustable number of independent wind plants. The model treats the generation level w of an individual wind farm as a random variable ranging from 0 to 1, where $w = W/W_{\textrm{cap}}$ represents the plant's output power $W$ normalized by its nameplate capacity $W_{\textrm{cap}}$. Transforming $10^7$ wind speed samples drawn from a Rayleigh distribution by the turbine power curve depicted in Figure~\ref{turbcurve}(a) (more details are in the Appendix to this chapter) yields a ``base'' sequence of $10^7$ independent samples of generation level w having the histogram shown in Figure~\ref{turbcurve}(b). For the chosen parameters the capacity factor or average generation level $\mu = 0.373$. The large count or spike in the histogram bin centered at $w = 0$ arises from ranges (i) and (iv) in Figure~\ref{turbcurve}(a) where wind speeds below cut-in and above cut-out give zero generation. Similarly the spike in the bin centered at $w = 1$ arises from range (iii) where wind speeds drive the turbines to rated capacity.

\begin{figure}[t]
\includegraphics[width=11.7cm]{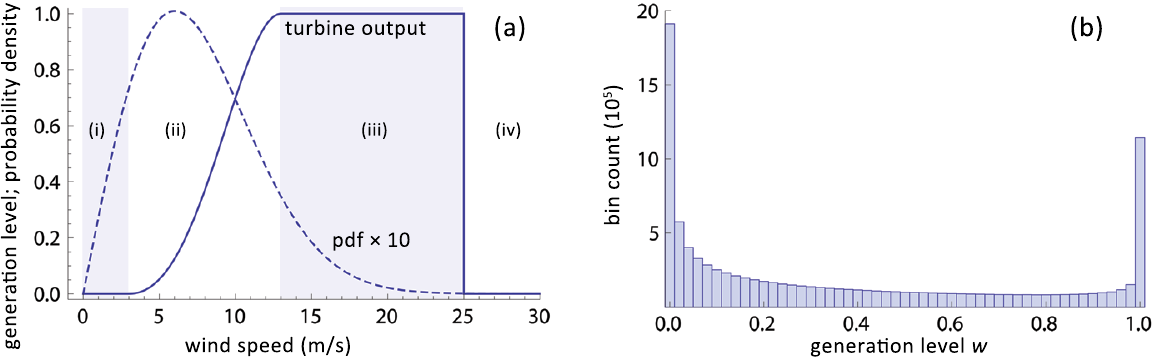}
\caption{(a) Rayleigh wind speed distribution with scale factor $s = 6$ m/s (dashed) and turbine power curve (solid). In range (i) below cut-in $v_1 = 3$ m/s generation is zero; in range (ii) between cut-in and rated speed $v_2 = 13$ m/s output depends on wind speed; in range (iii) between rated speed and cut-out $v_3 =25$ m/s output is constant at rated capacity, while in range (iv) beyond cut-out output falls again to zero. (b) Histogram of $10^7$ single-plant output power values.}
\label{turbcurve}
\end{figure}
While the histogram of Figure \ref{turbcurve}(b) is intended to be representative of the distribution of generation levels for a typical single turbine, we would expect the reduced variability of arrays of geographically diverse wind farms to be manifest in array histograms with rather different characteristics. We create a model of this behavior by supposing that the all the wind farms in a array are identical, each having the same generation-level probability distribution as the single turbine of Figure \ref{turbcurve}, but with generation levels independent of the other farms in the array. By taking ``diversity'' to the extreme of ``complete independence'' our model clearly shows the effects due to diversity. We treat the degree to which independence overstates the diversity of realistic geographically-distributed wind plants by later introducing the concept of effective sample size. 

To model the output of an array of $N$ independent wind plants we draw $N$ independent samples from the single-turbine distribution, add them together, and normalize them to the total array capacity. With the single-turbine levels already normalized, the array generation level is just the arithmetic average of the $N$ independent samples. For computational efficiency we create sequences for many different array sizes from a single $N = 1$ base sequence. Table~\ref{seqtbl} shows as an example a 5-element 3-sequence created from a 15-element base sequence. 
\begin{table}[t]
\includegraphics[width=11.7cm]{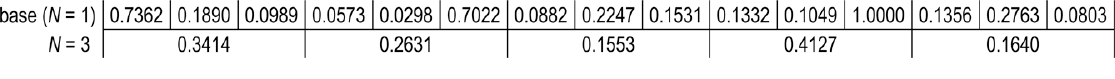}
\caption{Base sequence of random single-plant outputs ($N = 1$) and new 3-sequence of random outputs from an array of three independent plants ($N = 3$), generated by non-overlapping arithmetic averages of base-sequence triples.}
\label{seqtbl}
\end{table}

\section{Variability-reduction results}
\begin{figure}[b]
\includegraphics[width=11.7cm]{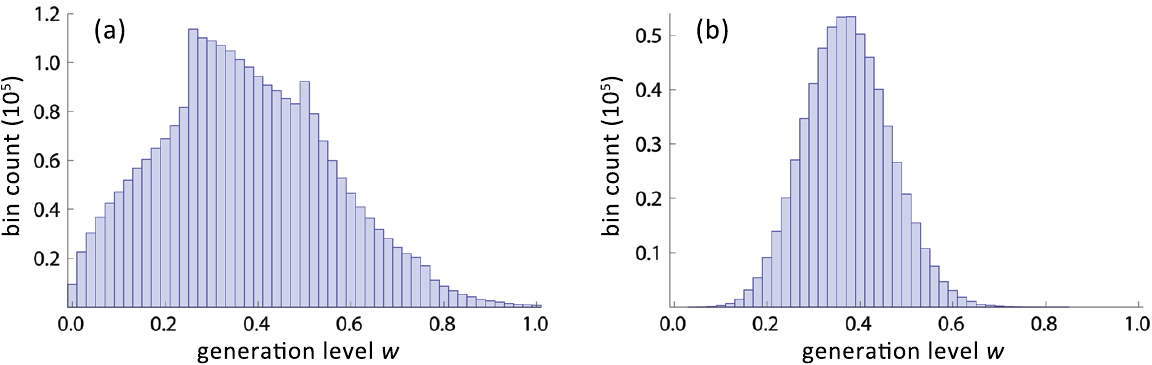}
\caption{Generation output histograms for arrays of (a) 4 and (b) 16 independent generators, with $2.5 \times 10^6$ and $6.25 \times 10^5$ array output power values, respectively.}
\label{histo}
\end{figure}

At small $N$ values the modeled array generation values cluster noticeably closer to the mean ($\mu = 0.373$ for all sequences here), as seen in Figure~\ref{histo}(a) for $N = 4$. Unlike the single-generator histogram of Figure \ref{turbcurve}(b) which has its most frequent values at the $w = 0$ and $w = 1$ extremes, the four-generator array histogram has a central mode, but still an appreciable fraction of zero-output generation levels. Increasing $N$ further drives the histograms toward the bell shape characteristic of the Gaussian or normal distribution, as expected according to the Central Limit Theorem and as seen in Figure~\ref{histo}(b) for $N = 16$.

\begin{figure}
\includegraphics[width=11.7cm]{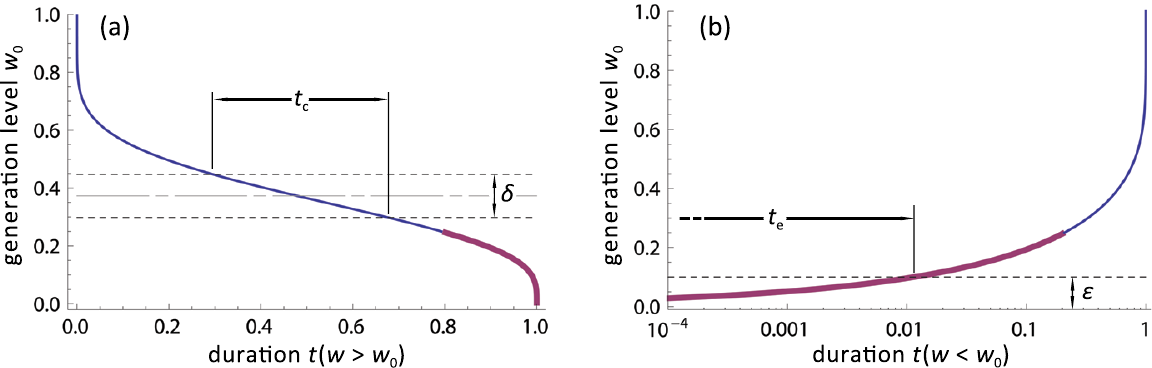}
\caption{Exemplary generation duration curves with definitions of two variability reduction benefits: (a) duration $t_c$ within range $\delta$ around mean; (b) duration $t_e$ within $\epsilon$ of $w = 0$. The tails of the curves covering same duration values in (a) and (b) are similarly colored to highlight the reversal of duration axis in (b).}
\label{durcurve}
\end{figure}

To quantify variability reduction we define two metrics shown in Figure~\ref{durcurve}. The amount of time $T$ the array of generators would spend in a particular state over a sufficiently long interval $T_{\textrm{tot}}$ gives a fractional ``duration'' $t = T/T_{\textrm{tot}}$ we associate with the frequency at which that particular state occurs in the $N$-sequence. Duration $t_c (\delta)$ defined in Figure~\ref{durcurve}(a) measures the tendency of generation level $w$ to lie within a range $\pm \delta/2$ of central mean $\mu$; here $\delta$ is 0.15 and the corresponding $t_c$ value is 0.38, meaning generation is within $\pm 7.5$\% of the mean 38\% of the time. Duration $t_e(\epsilon)$ defined in Figure~\ref{durcurve}(b) measures the tendency of generation to lie within a range $\epsilon$ above its low extreme level (zero); here $\epsilon$ is 0.10 and the corresponding $t_e$ value is 0.011, meaning that generation is less than 10\% about 1.1\% of the time. We have reversed the duration axis in Figure~\ref{durcurve}(b) relative that in \ref{durcurve}(a) so small values of $t_e$ can be clearly visualized on a conventional logarithmic axis. The two measures of reduced variability, larger $t_c$ and smaller $t_e$, depend differently on $N$, as we show.

\begin{figure}[b]
\includegraphics[width=11.7cm]{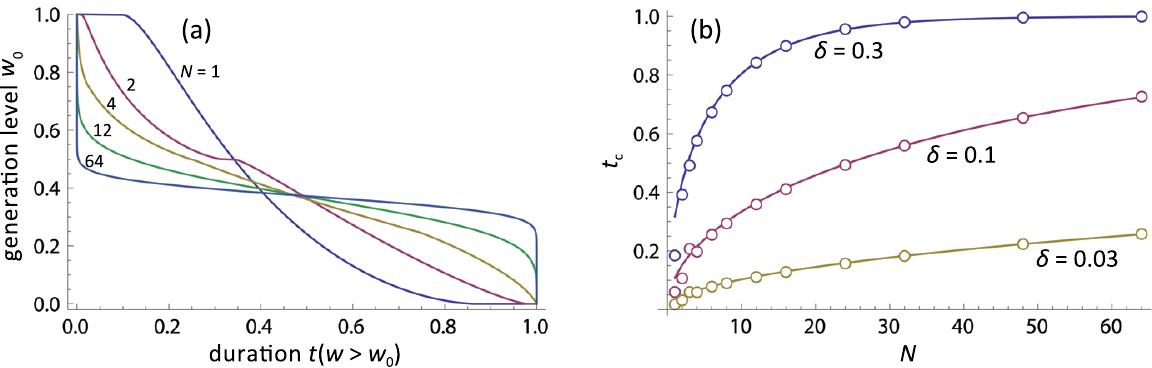}
\caption{Variability characteristics from Monte Carlo model for aggregations of $N$ identical independent generators. (a) Generation duration curves; (b) fraction of time $t_c$ that generation is within $\pm \delta /2$ of mean: Monte Carlo (symbols) with normal approximation (curves).}
\label{curvefam}
\end{figure}

First, consider the occurrence of near-mean generation levels. Figure~\ref{curvefam}(a) shows generation duration curves for arrays of different $N$. By ``$N$'' we do not mean individual wind turbines, or individual wind plants; rather, $N$ is the number of statistically independent wind generators in our Monte Carlo model. In a later section, we estimate the effective value of $N$ in various regions by using observed data. Increasing $N$ progressively flattens the curves around the mean, increasing duration $t_c \equiv t\left( \left| w - \mu \right| < \delta \right)$. Note the similarity of the $N = 1$ curve to that of BPA's 2009 output in Figure~\ref{emily}. 

Since the generation levels should become approximately normally distributed as $N$ increases we would expect duration $t_c$ to behave approximately as $t_c \approx 2\Upphi[\delta / \sigma_N] - 1$, where $\Upphi$ is the unit-normal cumulative distribution function (CDF). The variance is given by the Bienaym\'{e} formula as $\sigma_N^2 = \sigma_1^2/N$, where $\sigma_1^2 = 0.134$ is the single-plant variance. The normal approximation for $t_c$, plotted as solid curves in Figure~\ref{curvefam}(b), is quite accurate for $N$ larger than 5 or so. With near-normal behavior, $t_c$ increases in proportion to $\sqrt{N}$ until it begins to saturate at $t = 1$.

\begin{figure}[t]
\includegraphics[width=11.7cm]{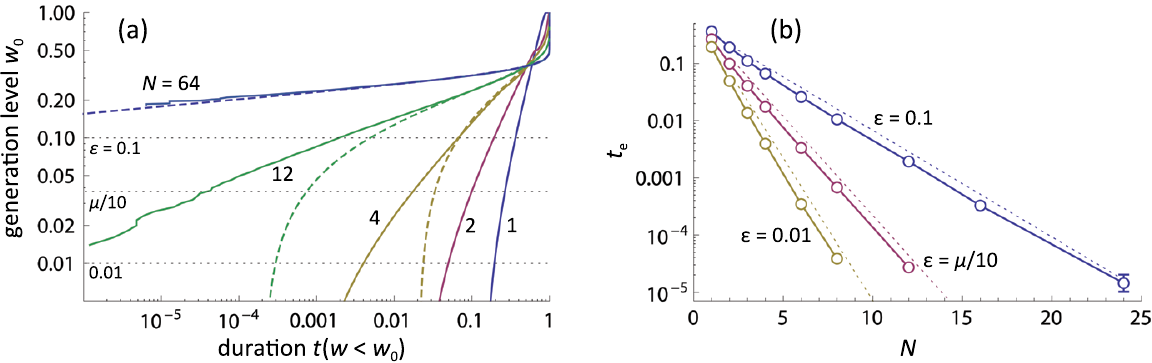}
\caption{Variability at low generation levels. (a) Reversed generation duration curves: Monte Carlo (solid curves) and normal approximation (dashed curves); (b) fraction of time $t_e$ that generation is less than $\epsilon$: Monte Carlo (symbols, solid curves), guide lines (dotted) of the form $be^{-\alpha N}$, with $\alpha = 0.43$, 0.80, and 1.15. Error bar on last $\epsilon = 0.1$ point spans the range of $t_e$ values from 10 Monte Carlo runs each with a different starting seed. }
\label{teesubee}
\end{figure}

The normal approximation, however, does not provide good estimates for the occurrence of near-zero generation levels, which we consider next. Figure~\ref{teesubee} shows reversed generation duration curves for arrays of the same $N$ values used in Figure~\ref{curvefam}. The increasingly square ``toe'' of the duration curve seen in Figure~\ref{curvefam}(a) correlates to the vanishing amounts of time spent at low generation levels observed in Figure~\ref{teesubee}(a). As can be seen by the dashed curves, the normal estimate $t \approx \Upphi \left[(w - \mu)/\sigma N \right]$ gives only a poor approximation to the Monte Carlo results for low generation levels. Variability metric $t_e \equiv t(w <  \epsilon)$ measuring the fraction of time that generation drops below threshold $\epsilon$ falls approximately exponentially with $N$ as shown in Figure~\ref{teesubee}(b), in accordance with the theory of large deviations \citep{lewis}. For our presumably typical turbine and wind parameters, and over the range of $\epsilon$ values down to at least 0.01, the rate of decline of $t_e$ with $N$ becomes larger the smaller $\epsilon$ gets. 

\begin{figure}[b]
\sidecaption
\includegraphics[width=6.1cm]{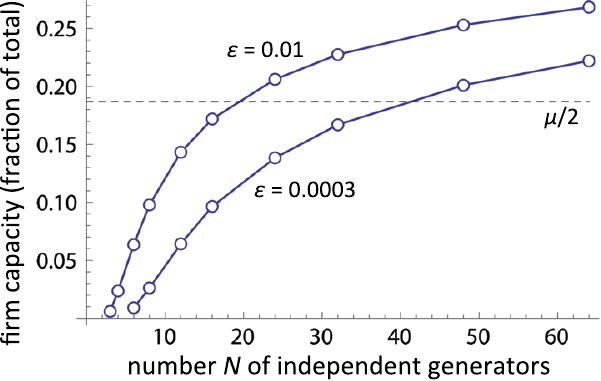}
\caption{Firm capacity modeled as a function of the number $N$ of independent generators for two allowed outage rates: 1\% and 0.03\% of the time.}
\label{firmcap}
\end{figure}
In evaluating electric-grid reliability a useful concept is that of firm capacity: capacity that is expected to be available all of the time except for infrequent outages. Figure~\ref{firmcap} translates the duration vs.\ capacity methods of Figure~\ref{teesubee} into the form of firm capacities. If an outage rate of 1\% were tolerable (as perhaps for an individual generator in a regional utility), a wind array could begin providing firm capacity with 3 or more independent farms. If outages were allowed only 0.03\% of the time, a wind array would begin providing firm capacity at $N = 6$. These model results ignore diurnal and seasonal variations in average wind speed which in fact have significant correlations with electric load that make meaningful capacity calculations more complex. 
\begin{figure}[b]
\includegraphics[width=11.7cm]{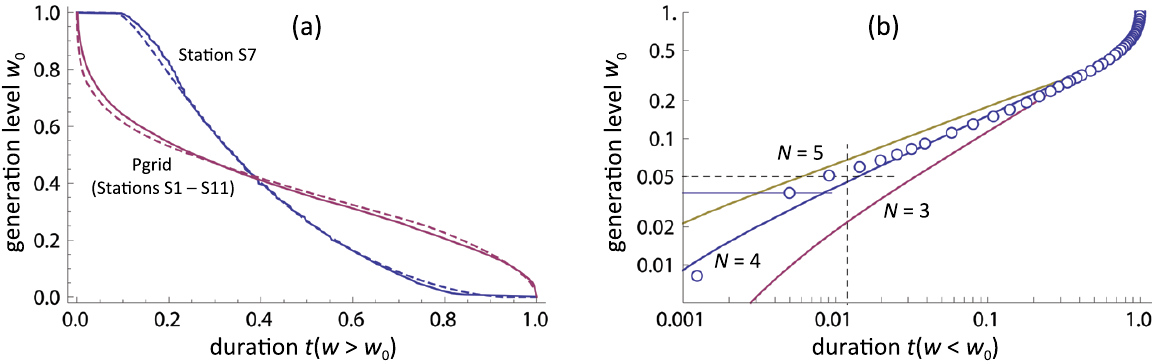}
\caption{Generation duration curves from plots in \cite{kempton}. (a) Single station S7 and all stations together (solid curves) compared to Monte Carlo model (dashed curves) for $N = 1$ and $N = 4$. (b) Reversed duration curves for $P_\textrm{grid}$(symbols) and Monte Carlo models for $N = 3$, 4, 5. Model results in (a) and (b) use $v_1 /s = 0.33$, $v_2 /s = 2.25$, $v_3 /s = 4.17$, $k/s = 2.89$.}
\label{kempton}
\end{figure}

\section{Comparison with observations}
Our model for the variability-reducing benefits arising from aggregation of independent wind generators compares well with observations based on real wind data as we show by examining results previously reported by \cite{kempton}. They created generation duration curves for 11 hypothetical offshore wind farms spread over the entire length of the U.S. Atlantic coast, and for the aggregate of all 11, called $P_\textrm{grid}$, using wind-speed data from meteorological buoys (see their Figure S3). In Figure~\ref{kempton}(a) we reproduce two of their twelve curves: one for single Station S7 (blue) that falls near the middle of the various capacity factors and one for $P_\textrm{grid}$ (red). Small adjustments to our turbine power curve parameters gave Monte Carlo model base sequence results ($N = 1$, blue dashed curve) approximating the single-station duration curve. Then, without further adjustments we compared model results for various array sizes to the $P_\textrm{grid}$ curve, selecting $N = 4$ as the best model match (red dashed). Figure~\ref{kempton}(b) shows the sensitivity of modeled low generation levels to choice of $N$: clearly $N = 3$ is too small and $N = 5$ is too large. (The error bar on the second point shows the width of Kempton's printed $P_\textrm{grid}$ curve from which we digitized the points; the dashed crosshairs slightly above the curve but below the digitization symbols mark the 1.2\% duration they explicitly report for $P_\textrm{grid} < 5$\%.) As nearly as can be determined from the published results, the variability-reducing benefits found by Kempton et al.\ for their geographically diverse Atlantic Transmission Grid are identical to the benefits of aggregating four statistically independent generators of similar characteristics.

\section{Discussion}
\begin{figure}[b]
\includegraphics[width=11.7cm]{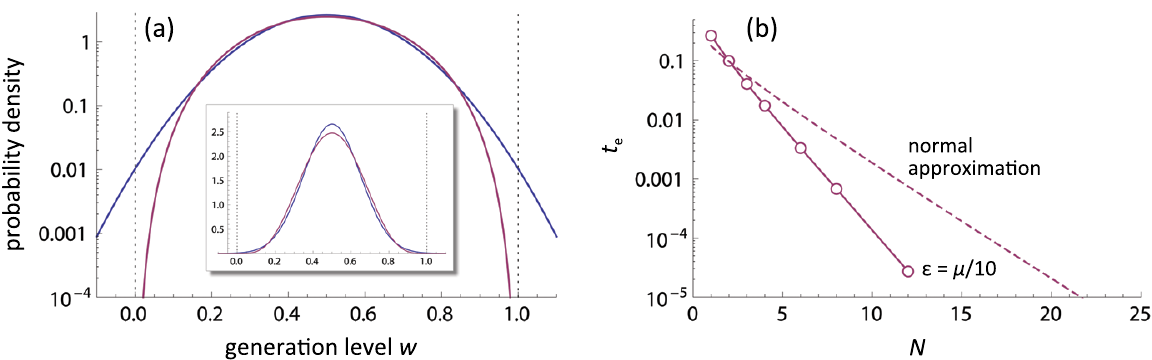}
\caption{Normal approximations. (a) Comparison of normal PDF (blue) with PDF of same mean (0.5) and standard deviation (0.15) but bounded to domain [0,1]: normal PDF tails extend significantly beyond domain boundaries (inset: linear axes). (b) Duration $t_e$ of generation less than 10\% of mean: Monte Carlo (symbols, solid curve), normal approximation (dashed curve).}
\label{normpdf}
\end{figure}

Our Monte Carlo model, along with well-defined quantitative metrics for variability reduction, provides a basis for understanding the benefits of geographically diverse wind arrays. Model calculations of the increase with $N$ of the duration $t_c$ of near-mean generation agree well with approximations made by assuming that aggregated generation levels are normally distributed. However, the same assumption does not suffice for estimating the duration $t_e$ of near-zero generation levels, for reasons that can be readily appreciated from the characteristics of the underlying PDF (probability density function) shown in Figure~\ref{normpdf}(a). While the domain of the normal PDF extends over $\left( - \infty, +\infty \right)$, any probability density representing generation normalized by nameplate capacity must necessarily equal zero outside of the interval [0,1], as shown by the red curve. Even though small, the area under the tails of the normal PDF outside the [0,1] interval is non-negligible, and will likely confound any attempted estimation of the probability (duration) of small generation levels. For example, Figure~\ref{normpdf} compares the Monte Carlo model results previously presented in Figure~\ref{teesubee}(b) for the duration of low generation levels to the corresponding normal approximation $t_e \approx \Upphi \left[(\epsilon - \mu)/\sigma  N \right]$. For $N = 12$ and $\epsilon = \mu /10 \approx 3.7$\%, the normal approximation overestimates the duration by a factor of about 30; it also estimates that 19 independent generators would be required to achieve a duration $t_e < 2.7 \times 10^{-5}$ when 12 would suffice. 

The non-zero temporal and geographical correlations of real wind speeds may raise concerns about the utility of our model, based as it is on independent random samples. Our model provides an upper bound to variability reductions achievable through geographic diversity because it is based on the assumption that every wind speed is uncorrelated with every other in space and time. However, real wind speeds are correlated with themselves in time and with speeds at sites nearby, reducing diversity. Real duration curves are usually measured over a time interval presumed long enough to characterize the range of typical system behavior. In this case temporal correlation effects in the real data should average out and the independent nature of the random samples in our model should not affect the realism of the results. 

With regard to spatial dependence, the close correspondence between our model results and the data reported by \cite{kempton} might be interpreted, however, in light of an effective sample size. For their buoy wind-speed data they found inter-site correlation falling as $\rho(r) = e^{- r / \ell}$ with $\ell = 430$ km. From the $L = 2600$~km end-to-end length of their array we would expect, by comparison with the diversity factor of \cite{farmer}, that the array would behave as though it comprised $N_{\textrm{eff}} \approx L/(2 \ell) = 2600/860 = 3.0$ independent generators. Comparing the average of the variances we extract from their 11 single-station duration curves, $\langle \sigma_1^2 \rangle = 0.132$, to that we extract from their $P_\textrm{grid}$ curve, $\sigma^2 = 0.042$, gives another estimate with a similar value: $N_{\textrm{eff}} \approx 0.132/0.042 = 3.1$. We find the agreement between these $N_{\textrm{eff}}$ values and the $N = 4$ of our closest-fitting model curve acceptable given, one, the difficulty in estimating correlation length $\ell$ from noisy meteorological data and, two, that we generated our Monte Carlo samples from a base distribution adjusted to match only a single ``representative'' site (S7), ignoring those other sites with lesser (or greater) $t_e$.

To gain further insight into $N_{\textrm{eff}}$ values, we examined generation data previously available from ERCOT, as analyzed by \cite{katz}, for 20 individual wind farms lying within a 450 km $\times$ 220~km region in Texas. Over the year 2008 the generation-level variances of the individual farms had a capacity-weighted average of 0.084 (range of 0.013 to 0.118), while the total generation had variance 0.055. Thus, comparing the variability of all wind generation in ERCOT to average of the variabilities of single farms leads to an apparent value for ERCOT of $N_{\textrm{eff}} = 1.5$. 

One might expect on the basis of size that a typical U.S. ISO would behave as though it comprised a number of independent generators lying between the $N_\textrm{eff} = 1.5$ we found for the 450-km ERCOT sub-region and the $N_\textrm{eff} = 4$  we found for Kempton's 2600-km $P_\textrm{grid}$. According to our modeled dependence of firm capacity on the number of independent generators such an ISO sits just on threshold of having enough diversity to yield the beginning of any firm wind capacity. Utilizing transmission lines to connect several such regions could give non-negligible firm capacity, although firming with conventional reserve generation would likely be more cost effective.

However, before taking the effective sample size concept too seriously, we need a better understanding of the nature of correlation vs. separation for real wind generators. In particular, if the infrequent outages responsible for the outages at low rates are also short then the appropriate correlation length  may be smaller than it is for longer fluctuations \citep{ernst}, resulting in a larger $N_{\textrm{eff}}$ than might be expected from the ratio of variances. Moreover, most work to date on wind-speed correlation vs. site separation characterizes correlation only by its single coefficient, with no assurance that this number equally represents behaviors at low and average wind speed levels. 

\section{Summary}
This chapter presents a simple model for the benefits of geographic diversity of wind generation based on arrays comprising a selectable number $N$ of statistically independent wind generators. The $N$ statistically independent generators in the model correspond to a greater number of partially correlated generators in a real array. We emphasize that $N$ is not the number of actual wind turbines or wind plants. For electric power generated by $N$ statistically independent wind plants, the duration of near-mean generation grows at a rate initially proportional to $\sqrt{N}$ that then declines only as the generation level approaches being within the near-mean bounds 100\% of the time. The incremental or marginal benefit from each added statistically independent generator declines as $N^{-1/2}$. Duration $t_e$ of near-zero generation decreases approximately as $e^{-\alpha N}$, where rate $\alpha$ becomes larger as the generation threshold gets closer to zero. Log$[t_e]$ never saturates, and has constant marginal improvement from each incremental independent generator. 

The concept of effective sample size helps differentiate between the effect of increasing generator number and increasing geographic area. An array of real generators within a region of linear size comparable to correlation length  has high inter-generator correlation and never gives $N_{\textrm{eff}}$ much larger than 1, yielding a variability reduction that saturates rapidly with generator number, as shown systematically by \cite{hasche}. On the other hand, to the extent that inter-generator correlation falls off exponentially with separation, there is no limiting distance or region size beyond which further expansion fails to produce growth in achievable $N_{\textrm{eff}}$ and substantial additional variability reduction.

Further work on the nature of wind correlation is needed to determine whether sufficiently large $N$ values are feasible within continental-scale geographic regions to provide substantial firm capacity.

\section{Appendix: turbine power curve}
Our Monte Carlo model converts each randomly generated wind speed sample $v_i$ to a generation level $w_i$ using the following turbine power function:
\[  w_i = \left\{
\begin{array}{ll}
      0 & v_i < v_1 \; \textrm{or} \; v_i \geq v_3\\
      \sin \left[ \frac{\pi}{2} \left( \frac{v_i - v_1}{v_2 - v_1} + \frac{(v_i - v_1)(v_i - v_2)}{k^2} \right) \right] & v_1 \leq v_i < v_2 \\
      1 & v_2 \leq v_i < v_3 
\end{array} 
\right. \]
with $v_1 = 3$ m/s, $v_2 = 13$ m/s, $v_3 = 25$ m/s, and $k = \sqrt{300}$ m/s, unless otherwise noted. The function provides a continuous, invertible and differentiable approximation to the manufacturer's data for the GE 1.5 MW turbine.




\begin{thebibliography}{}

\bibitem[Archer and Jacobson(2003)]{archer}
Archer, C. L. and M. Z. Jacobson. 2003. ``Spatial and temporal distributions of U.S. winds and wind power at 80 m derived from measurements.'' \textit{Journal of Geophysical Research} 108:4289. \doi{10.1029/2002jd002076}.

\bibitem[Archer and Jacobson(2007)]{archer2}
Archer, C. L. and M. Z. Jacobson. 2007. ``Supplying baseload power and reducing transmission requirements by interconnecting wind farms.'' \textit{Journal of Applied Meteorology and Climatology} 46:1701--1717. \doi{10.1175/2007JAMC1538.1}.

\bibitem[Beyer, Luther, and Steinberger-Willms(1993)]{beyer}
Beyer, H. G., J. Luther, and R. Steinberger-Willms. 1993. ``Power fluctuations in spatially dispersed wind turbine systems.'' \textit{Solar Energy} 50:297--305. \doi{10.1016/0038-092X(93)90025-J}.

\bibitem[Carlin and Haslett(1982)]{carlin}
Carlin, J. and J. Haslett. 1982. ``The probability distribution of wind power from a dispersed array of wind turbine generators.'' \textit{Journal of Applied Meteorology} 21:303--313. \doi{10.1175/1520-0450(1982)021<0303:tpdowp>2.0.co;2}.

\bibitem[Dvorak et al.(2012)]{dvorak}
Dvorak, M. J., E. D. Stoutenburg, Cristina L. Archer,  Willet Kempton, Mark Z. Jacobson. 2012. ``Where is the ideal location for a U.S. East Coast offshore grid?'' \textit{Geophysical Research Letters} 39:L06804.  \doi{10.1029/2011gl050659}.

\bibitem[Ernst, Wand, and Kirby(1999)]{ernst}
Ernst, B., Y.-H. Wan, Brendan Kirby. 1999. \textit{Short-Term Power Fluctuation of Wind Turbines: Analyzing Data from the German 250-MW Measurement Program from the Ancillary Services Viewpoint.} National Renewable Energy Laboratory, Report NREL/CP-500-26722. http://www.osti.gov/scitech/servlets/purl/9826.

\bibitem[Farmer, Newman, and Ashmole(1980)]{farmer}
Farmer, E. D., V. G. Newman, P. H. Ashmole. 1980. ``Economic and operational implications of a complex of wind-driven generators on a power system.'' \textit{IEE Proceedings Part A} 127:289--295. \doi{10.1049/ip-a-1:19800046}.

\bibitem[Fertig et al.(2012)]{fertig}
Fertig, Emily, Jay Apt, Paulina Jaramillo, Warren
Katzenstein. 2012. ``The effect of long-distance interconnection on wind power variability.'' \textit{Environmental Research Letters} 7:034017. \doi{10.1088/1748-9326/7/3/034017}.

\bibitem[Fisher et al.(2013)]{fisher}
Fisher, Samuel M., Justin T. Schoof, Christopher L. Lant, 
Matthew D. Therrell. 2013. ``The effects of geographical distribution on the reliability of wind energy.'' \textit{Applied Geography} 40:83--89. \doi{10.1016/j.apgeog.2013.01.010}.

\bibitem[Hasche(2010)]{hasche}
Hasche, Berhard. 2010 ``General statistics of geographically dispersed wind power.'' \textit{Wind Energy} 13:773--784. \doi{10.1002/we.397}.

\bibitem[Holttinen(2005)]{holt}
Holttinen, Hannele. 2005. ``Hourly wind power variations in the Nordic countries.'' \textit{Wind Energy} 8:173--195. \doi{10.1002/we.144}.

\bibitem[Huang, Lu, and McElroy(2014)]{huang}
Huang, Junling, Xi Lu, Michael B. McElroy. 2014. ``Meteorologically defined limits to reduction in the variability of outputs from a coupled wind farm system in the Central U.S.'' \textit{Renewable Energy} 62:331--340. \doi{10.1016/j.renene.2013.07.022}.

\bibitem[Justus and Mikhail(1978)]{justus}
Justus, C. G. and A. S. Mikhail. 1978. \textit{Energy statistics for large wind turbine arrays.} Georgia Institute of Technology, http://hdl.handle.net/1853/40482.

\bibitem[Kahn(1979)]{kahn}
Kahn, E. 1979. ``The reliability of distributed wind generators.'' \textit{Electric Power Systems Research} 2:1--14.  \doi{10.1016/0378-7796(79)90021-X}.

\bibitem[Katzenstein, Fertig, and Apt(2010)]{katz}
Katzenstein, Warren, Emily Fertig, and Jay Apt. 2010. ``The variability of interconnected wind plants.'' \textit{Energy Policy} 38:4400--4410. \doi{10.1016/j.enpol.2010.03.069}.

\bibitem[Kempton et al.(2010)]{kempton}
Kempton, Willett, Felipe M. Pimenta, Dana E. Veron, Brian A. Colle. 2010. ``Electric power from offshore wind via synoptic-scale interconnection.'' \textit{Proceedings of the National Academy of Sciences} 107:7240--7245.  \doi{10.1073/pnas.0909075107}.

\bibitem[Lewis and Russell(1997)]{lewis}
Lewis, John T. and Raymond Russell. 1997. \textit{An introduction to large deviations for teletraffic engineers.}  Computer Laboratory, University of Cambridge, www.cl.cam.ac.uk/research/srg/netos/measure/tutorial/rev-tutorial.ps.gz013.

\bibitem[Louie(2014)]{louie}
Louie, Henry. 2014. ``Correlation and statistical characteristics of aggregate wind power in large transcontinental systems.'' \textit{Wind Energy} 17:793--810. \doi{10.1002/we.1597}.

\bibitem[McNerney and Richardson(1992)]{mcnerney}
McNerney, G. and R. Richardson 1992. ``The statistical smoothing of power delivered to utilities by multiple wind turbines.'' \textit{IEEE Transactions on Energy Conversion} 7:644--647, \doi{10.1109/60.182646}.

\bibitem[Molly(1977)]{molly}
Molly, J. P. 1977. ``Balancing power supply from wind energy converting systems.'' \textit{Wind Engineering} 1:57--66.

\bibitem[Nanahara et al.(2004)]{nan}
Nanahara, Toshiya, Masahiro Asari, Tsutomu Maejima, Takamitsu Sato, Koji Yamaguchi, Masaaki Shibata. 2004. ``Smoothing effects of distributed wind turbines. Part 2. Coherence among power output of distant wind turbines.'' Wind Energy 7:75--85, \doi{10.1002/we.108}.

\bibitem[Tarroja et al.(2011)]{tarroja}
Tarroja, Brian, Fabian Mueller, Joshua D. Eichman, Jack Brouwer, Scott Samuelsen. 2011. ``Spatial and temporal analysis of electric wind generation intermittency and dynamics.'' \textit{Renewable Energy} 36:3424--3432. \doi{10.1016/j.renene.2011.05.022}.

\bibitem[Thomas(1945)]{thomas}
Thomas, P. H. (1945). \textit{Electric Power from the Wind: A Survey.} United States Federal Power Commission.

\end{thebibliography}
\end{document}